# FMI Compliant Approach to Investigate the Impact of Communication to Islanded Microgrid Secondary Control

Tung Lam NGUYEN[1,2], Quoc-Tuan TRAN[2*], Raphael CAIRE[1], Yvon BESANGER[1*], Tran The HOANG[1], Van Hoa NGUYEN[1]
[1] University Grenoble Alpes, G2Elab, F-38000 Grenoble, France
CNRS, G2Elab, F-38000 Grenoble, France
[2] CEA-INES, Le Bourget-du-lac, France
*Senior Member, IEEE
tung-lam.nguyen @g2elab.grenoble-inp.fr

*Abstract*—In multi-master islanded microgrids, the inverter controllers need to share the signals and to coordinate, in either centralized or distributed way, in order to operate properly and to assure a good functionality of the grid. The central controller is used in centralized strategy. In distributed control, Multi-agent system (MAS) is considered to be a suitable solution for coordination of such system. However the latency and disturbance of the network may disturb the communication from central controller to local controllers or among agents or and negatively influence the grid operation. As a consequence, communication aspects need to be properly addressed during the control design and assessment. In this paper, we propose a holistic approach with co-simulation using Functional Mockup Interface (FMI) standard to validate the microgrid control system taking into account the communication network. A use-case of islanded microgrid frequency secondary control with MAS under consensus algorithm is implemented to demonstrate the impact of communication and to illustrate the proposed holistic approach.

*Index Terms*—microgrid, hierarchical control, function mockup interface, multi-agent system, communication network,

## I. INTRODUCTION

Microgrids (MGs) are essential blocks to constitute and are considered to be one of the major changes required in the development of the power system. In general, a MG consists of a cluster of distributed generators (DGs), loads, energy storage systems and other equipment, which can operate in islanded mode or grid-connected, and can seamlessly transfer between these two modes [1]. In the islanded mode, due to the lack of the bulk grid as a reference, the microgrid has to keep its stable operation by coordinating all the elements of its own.

The control structure of microgrid is typically hierarchical and divided into three levels including primary, secondary and tertiary. The primary control level only uses the local information to response quickly to the change of the system. On the contrary, the secondary and tertiary control level requires the remote information. The information is global or adjacent depends on the microgrid controlled in centralized or distributed fashion respectively. A centralized control microgrid requires a central controller to send the appropriate signals to every local controller. Meanwhile, the microgrid with distributed control strategy only needs the communication between neighbor local controllers and the dependent on the central controller could be ignored. The multi-agent system method is widely used in the latter control strategy. In either case, the power system control requires signal from other spots to operate correctly and any disturbances of communication network (i.e. latency, packet loss) would negatively influence its functionality and subsequently affect the performance and even stability of the system [2]. Therefore, communication aspects need to be taken into consideration in microgrid control assessment[3].

Currently, the impacts of communication on control in a microgrid are studied in several approaches. A typical approach is to use co-simulation method [4]. In general, one simulation environment only supports only one specific domain (PowerFactory, Matlab/Simulink, Powerworld,… in power system domain or ns-2, ns-3, OMNET++,… in communication domain). Therefore, the comprehensive investigation in a hybrid system needs the combination of two or more simulators. However, the synchronization of the signals exchanged between simulators and the appropriate application programming interface (API) are still the challenges of the co-simulation method. The another approach is to design a system with the physical controllers and the real network to transmit data [5]. The grid of the system is run in real-time and is controlled by hardware controllers. By setting up the real communication network, the data is transferred in a natural way and the behavior could be close to the system in reality. However, due to the limitation of area of testing laboratories, the distances between the controllers could not be reflected exactly.

In this paper, we firstly propose a co-simulation based method to validate the secondary control strategies in island microgrid taking into account the impact of the communication network. In this method, we use the standard Function Mockup Interface (FMI) to transfer the communication emulation into the power system simulation environment. Secondly, we apply the proposed method to check the performance of a multi-master islanded microgrid in various scenarios of the communication network. A multi-agent system platform is also developed in the distributed control circumstance.

The rest of the paper is organized as follows. Section II introduces the primary and secondary control in MG. The introduction of FMI and the proposed method is presented in Section III. In section IV, the experimental setup and results are shown to validate the proposed method. Section VI concludes the paper and outlines possible future directions.

## II. SECONDARY CONTROL IN ISLANDED MICROGRID

The microgrid control with inverter-based DGs mainly refers to the inverter control due to the fact that it is commonly interfaced with the prime energy of each DG via a power electronic inverter. Therefore, the frequency control in microgrid is regulated by the coordination of inverter controllers.

The primary control, or local control, adjusts the frequency and amplitude of voltage reference provided to the inner control loop of voltage source inverter. Droop control method is used to control power sharing between DGs in MG without communication. The droop equation presenting the relationship between frequency and active power is:

$$f = f_0 - k_P(P - P_0) \quad (1)$$

where $f_0$ is rated frequency of grid voltage and $P_0$ is the normal value of real power. $f$ is the actual measured value of frequency when the DG is supplying real power of $P$. $k_P$ is the droop coefficient.

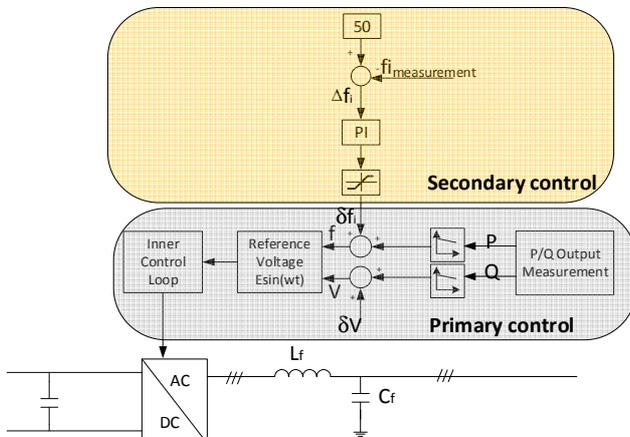

Figure 1. The inverter controller

The primary control maintains the voltage and frequency stability in MG. However, the frequency of microgrid after the primary control deviates from the rated frequency. This deviation will be eliminated in secondary control level. The steady-state error is measured from the grid and compensated by a PI controller. In particular, the secondary control is computed as

$$\delta_f = K_P \Delta f + K_i \int \Delta f \quad (2)$$

where $K_P$ and $K_i$ are the control parameter of PI controller, $\Delta f$ is the measured microgrid frequency deviation and $\delta_f$ is the secondary control signal sent to primary control level and then added to the correction given by each loop controller in Equation 1.

$$f = f_0 - k_P(P - P_0) + \delta_f \quad (3)$$

The secondary control level method could be divided into centralized and distributed control.

Centralized control schemes for power systems are common. In the traditional centralized control algorithms, a central controller is required to collect all information from local controllers and process a large amount of data, which suffers from computation stress and single-point-failure. In the case of the secondary control, the complement frequency is sent to all primary controllers of DGs. Such an approach has the advantage of having a global overview of the whole system. However, such a system is not easily extendable, is far from being computationally scalable and has the vulnerability of a single point of failure.

Alternatively, the distributed control with a spare communication network does not need a central controller and each unit is controlled by its local and neighbor control system. The distinct feature of the distributed approach is that the information involved in the control algorithm is not global, but adjacent for any given unit. Also, the length of the communication links is often shorter, which offers better and more reliable latency. Moreover, the risk of overall system failure can be reduced, because the system does not depend on a sole central controller.

The algorithms of centralized and distributed secondary control will be discussed more detail in the next section with a specific test case of a microgrid.

Due to the necessary of transferring data in both cases of secondary control, which is global in term of centralized control or adjacent in term of distributed control, the communication network could affect the control operation and maybe the stability of the microgrid. The communication network, in reality, is imperfection and constraint. Therefore, the evaluation of a control system needs an extra consideration of the operation of the transmission network.

## III. FMI COMPLIANT CO-SIMULATION APPROACH TO INVESTIGATE THE IMPACT OF COMMUNICATION SYSTEM

### A. Co-simulation of Power system and communication network

While communication technology increases rapidly its penetration in the power system, there are a very limited number of methodologies and tools that allows the operators to take into account both domains in a holistic manner. In the domain of smart grid nowadays, the co-simulation approach is often used to couple a power system simulator and a communication simulator. Co-simulation framework allows in general the joint and simultaneous simulation of models developed with different tools, in which the intermediate results are exchanged during simulation execution. The works on co-simulation of hybrid system Power/Com can be found in [6][7]. Generally, we acknowledge two structures of co-simulation:

- Ad-hoc co-simulation: coupling directly one power system simulator and one communication network simulator.
- Co-simulation with the Master algorithm: a master algorithm (e.g. HLA [8]) or a co-simulation framework (e.g. Mosaik [9], Ptolemy [10]) will orchestrate the process. This master algorithm is responsible for synchronizing different timelines of involved simulators and for directing the information exchange among simulator's inputs/outputs.

Co-simulation is still a difficult method for the electrical engineering community due to the necessity of synchronizing both simulation tools properly at runtime. While power system simulation is normally continuous with the possibility of event detection associated to value crossing a certain threshold; communication network simulation is based on discrete events whose occurrence usually stochastically distributed with respect to time. The simulator provides an event scheduler to record current system time and process the events in an event list. Moreover, the existing simulation tools offer limited options of adequate Application Programming Interface (API) for external coupling.

*B. Proposition to integrate communication emulation into the power system environment*

To investigate the impact of communication network to power system, we propose in this paper a method using Function Mockup Interface (FMI)[1] standard which allows interoperability and reusability of the models in co-simulation frameworks.

FMI is a standard designed to provide a unified model execution interface for dynamic system models between modeling tools and simulation tools. The idea is that tools generate and exchange models that adhere to the FMI specification. Such models are called Functional Mock-up Units (FMUs). Since its release, FMI has received a significant attention from both tool vendors and users. According to the information on the official website, there are currently over 101 tools that support or plan to support FMI. There is a real and pressing need to be able to export and import dynamic system models between existing tools, and also to be able to develop custom simulation environments.

Based on the FMI standard, we build a communication FMU to simulate the latency and to emulate the according to packet loss of the considered communication. The latency can be calculated based on one-sided transmission or round trip time. Due to the nature of our systems of interest, we consider only one-sided transmission latency or the time for the signal from sender to reach the destination. The model allows us to emulate the communication between two points in a network. In our problem, latency is defined as the time interval between the emission of first bit from the sender and the arrival of last bit of a signal in the destination. In order to represent the stochasticity, a white noise is added to the latency block formula. In reality, the communication is done via several intermediate servers with different protocols and technologies in-between. The communication FMU takes that information on network topology into account. A protocol library is available to provide the message transfer module with information about the employed protocols.

A general configuration of a communication FMU can then be illustrated in Figure 2. In this version, the transmission speed and data rate need to be manually defined. In future development, this information could be automatically acquired from the protocol library. The Communication FMU is then integrated into the considered power system to evaluate the impact of communication network to system performance.

In the next section, we apply the proposed method to an islanded microgrid secondary control, using centralized and MAS approaches.

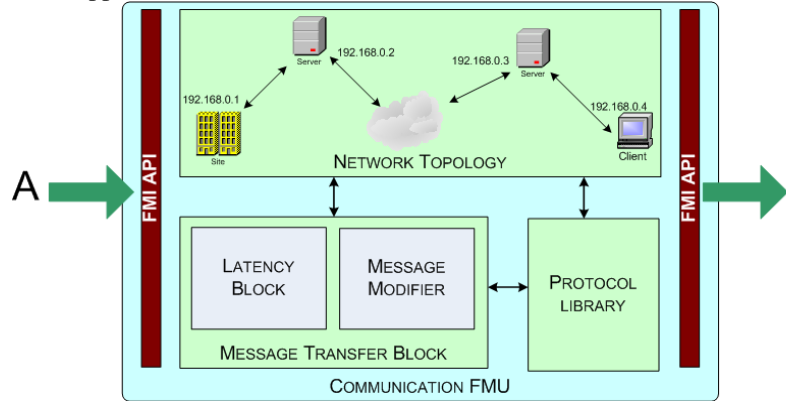

Figure 2. General elements of a communication FMU

IV. IMPLEMENTATION AND RESULTS

*A. Test-case description and set up*

In this paper, we apply the proposed holistic approach to a test-case of operating a micro-grid in island mode. The MG includes five distributed generators (DGs) supplying power for two loads. For the sake of simplicity, the primary sources of DGs are supposed to be an ideal DC source. Each DG is controlled by a local controller as Figure 1. The controllers send appropriate pulses signal to inverters to keep the frequency and voltage amplitude of the grid in reference values. The DGs, which are controlled in the voltage/frequency mode (inverter is the grid-forming inverter), operate in parallel so coordination is required. The system is studied from the time at 60 seconds when the Load 2 is tripped, i.e. the total load is diminished and the control has the responsibility to return the frequency back to normal condition.

We will validate the control system of the described microgrid taking into account the communication network in both cases of the secondary control - centralized or distributed. In the centralized approach as *Figure 3.a*, we use a microgrid central controller (MGCC) to diffuse the value of complement frequency to local controllers. Meanwhile, in the distributed approach as *Figure 3.b*, the Multi-agent system (MAS) with the consensus algorithm is used to exchange information. We investigate in four cases of the network:
- The first case is the network with ideal conditions; the influence of communication network could be ignored.
- In the three remaining cases, the data transmission is considered by adding the FMU communication block. The network scenarios 1, 2 and 3 are sorted by the decay level of quality, i.e. growing latency (cf. Table 1).

The grid in this paper is simulated in Matlab/Simulink by using the SimPowerSystems toolbox. In the library of Malab/Simulink, there are blocks which can use to express the delay in transmission of data. However, the variation of latency must be defined by the user. The time of delay, in fact, is not always easy to estimate. It varies in a range in a natural way depending on the properties of the network such as the transmitted distance, the protocol, the size of data, etc. The emulation of the network, therefore, need a more precise module. In our work, we developed the module in the FMI standard and integrate to the control loop of the model using the Pilot Support Package (PSP). Instead of transferring directly, the data is sent and received through the FMU block. This could express the delay in the transmission between two distinct points. The transmission is therefore emulated in the nearly natural way and could be applied to investigate any systems which need the broadcast of data.

---

[1] https://www.fmi-standard.org/

*1) Centralized control*

Figure 3.a illustrates the islanded microgrid controlled in centralized strategy. The MGCC is assumed to be located at a point which has distances to the local controllers as Table 1. The MGCC is in charge of the secondary control. It has the communication links with all local controllers. The measurement frequency at one point in the microgrid is sent to the MGCC. The complement frequency is then calculated at the secondary level and diffused to the primary control units which are put at the local sides of the DGs.

TABLE 1. THE DISTANCES OF COMMUNICATION FROM MGCC TO CONTROLLERS

| MGCC-Controller 1 (c-1) | MGCC-Controller 2 (c-2) | MGCC-Controller 3 (c-3) | MGCC-Controller 4 (c-4) | MGCC-Controller 5 (c-5) |
|---|---|---|---|---|
| 4 km | 6 km | 8 km | 2 km | 5 km |

We investigate the system in four described cases where the distances remain the same as prescribed in Table 1. However, the other properties of the network (i.e. Data rate, serialization speed and bandwidth) changed and caused the latency in different ranges of values (Figure 4). We use the box-and-whisker plot to illustrate the distribution of the datasets of the delay time with 1000 random samples in each case.

The frequency of the system is demonstrated in Figure 5. In all cases, the control system brings the frequency back to the reference values after the variation happens in the microgrid. This verified that the designed control guarantee the stability of the operation of the grid with ideal communication and with considered communication network scenarios.

*2) Distributed control*

We propose a structure using a multi-agent system that leverages the consensus algorithm in the context of microgrid distributed control. The multi-agent system layer is added on top of the control layer. The agents in the multi-agent system are put at locations of DG units and take the responsibility of processing and exchanging information. The difference between two control strategies is that the distributed control fashion requires the information from neighbor agents. Therefore, the inter-unit communication with the distances described in Table 2 could also affect to the control system. We also investigated the system in the four cases of communication network. The latency in inter-agent transmission in three cases of the communication network is illustrated in Figure 6.

TABLE 2. THE DISTANCES OF COMMUNICATION BETWEEN AGENTS

| Agent 1-Agent 4 (1-4, 4-1) | Agent 2-Agent 3 (2-3, 3-2) | Agent 3-Agent 4 (3-4, 4-3) | Agent 3-Agent 5 (3-5, 5-3) |
|---|---|---|---|
| 4 km | 6 km | 8 km | 2 km |

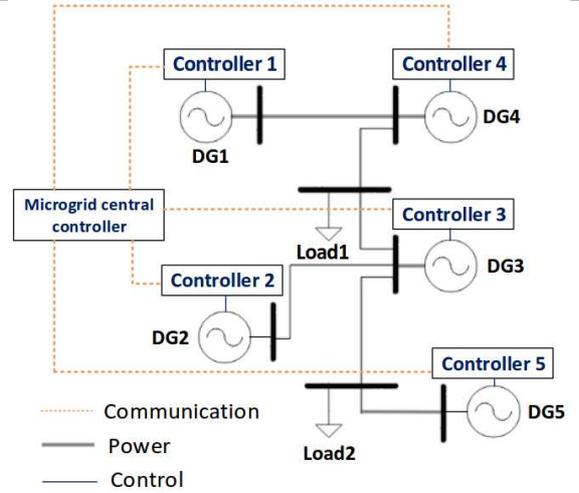

a) Centralized control

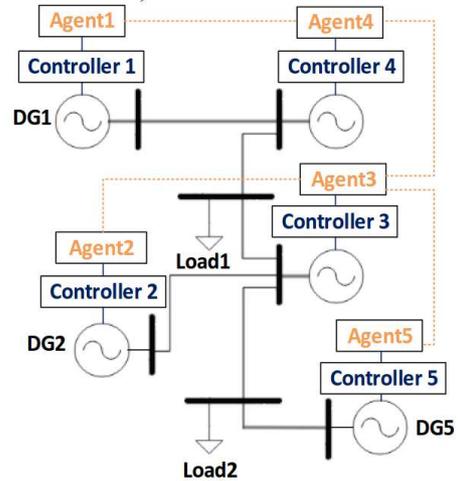

b) Distributed control

Figure 3. The test-case configuration

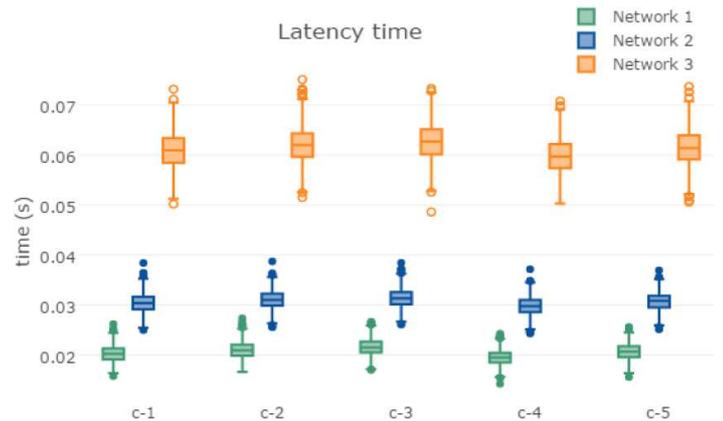

Figure 4. The latency time when transmitting data from the MGCC to the controllers in the three test cases

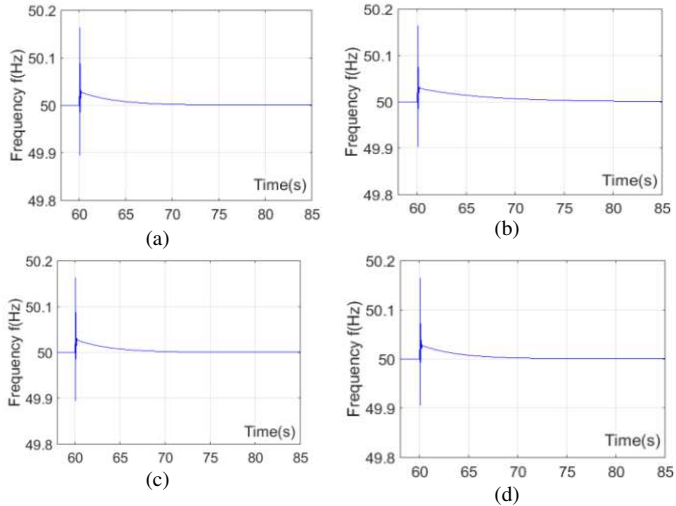

Figure 5. The frequency when tripping load 2
a) without communication network, b) with latency test case 1, c) with latency test case 2, d) with latency test case 3

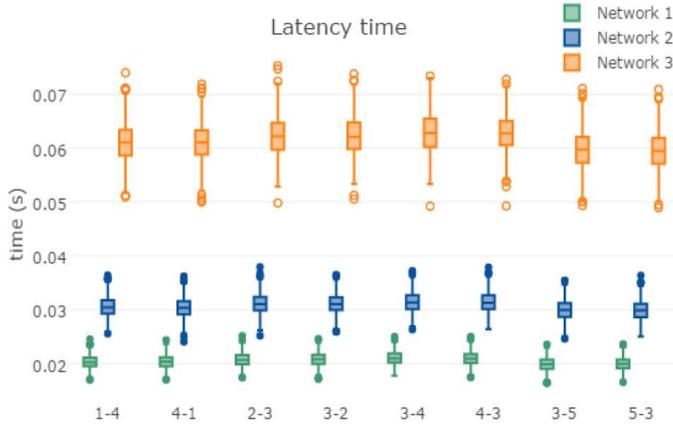

Figure 6. The inter-agent latency time in the three test cases

Agent layer takes the responsibility alike the microgrid central controller. The agent layer sends signals to control layer to bring the frequency back to the reference value. However, the frequency at DGs oscillates in the transient period after a variation in the microgrid. Due to the grid operates in the multi-master control mode, the requirement is that the signals sent to all local controllers at the almost same time and those signals have the same value to ensure the property of the control operation. To reach these conditions, we applied the average consensus algorithm. The result is that the local controllers could get simultaneously the average of all instantaneous frequency deviations at the output of DGs.

In the literature, the agent-based distributed control in a microgrid is usually implemented by connecting with a multi-agent platform [11] such as JADE, ZEUS, aiomas, etc. The interface which connects the simulation of the grid system to the platforms could be a barrier in studying. In this paper, we built the agent system platform right in the simulation of the grid. This platform is created based on the idea of integrating the communication FMU into the power simulation environment. Figure 7 presents the agent designed in Matlab/Simulink.

More details of the application of the consensus algorithm could be found in [12]. The main idea of the progress in each agent is that it gets the instant measurement from the system and returns the mean value to the controller. The progress is calculated in an iterative way. Suppose that $N^k$ denotes the set of the agent has the communication with agent $k$, $a$ denotes the matrix calculated following the Metropolis rule [13], $f_i^j$ denotes the values of calculated frequency at iterator $i$ of agent $j$. The pseudo code of the process inside agent $k$ is described as following:

1) The iterator $i = 1$
2) The agent receives the instantaneous value of the frequency at output of the corresponding DG
$$f_i^k = f_{meas}$$
3) The agent sends the signals including the current iterator and agent value to the neighbors
4) The agent receives the signals from the neighbors
5) *If* the agent receives signal at iteration $i$ from every single agent in its neighbor set *then*
$$i = i + 1$$
$$f_{i+1}^k = a^{kk} * f_i^k + \sum_{j \in N^k} a^{kj} * f_i^j$$

   *Else* move to Step 4
6) *If* $i = n_{consensus}$ *then* move to Step 6
   *Else* move to Step 3
7) Send the agent value to the corresponding controller
   Move to Step 1

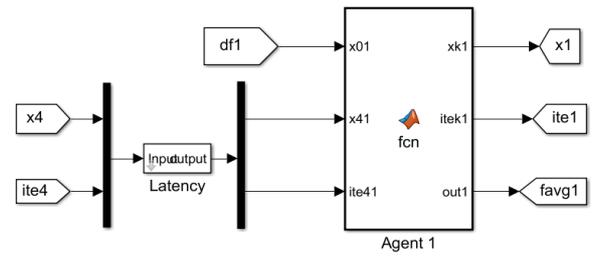

Figure 7. An agent in Matlab/Simulink

The controller gets signals when the corresponding agent reaches the consensus or average value at Step 7. This means that the frequency at the input of the controller is updated after $n_{consensus}$ iterators. The delay in transmission, therefore, depends on not only the latency when sending and receiving data between agents but also the processing time inside agents.

The frequency of the system is demonstrated in Figure 8. In the test case without network latency or with the high quality network (Network 1 and Network 2), after a short variation due to the disturbance in the microgrid, the frequency was restored correctly. Nevertheless, in Network 3 system, the frequency was no longer stable. In this case, the duration of transmitting data is much more than the other cases and therefore it increases significantly the time of consensus process in multi-agent system. As a consequence, the Proportional-Integral (PI) control in the secondary level works improperly issuing wrong or late decision and the frequency could not return to the static rated value. Therefore the control system needs to be adjusted to fulfill the stability requirement of the system. In particular, we tuned parameters of the PI block by increasing the Integral factor $K_i$ in Equation 2. The frequency of the modified system is shown in Figure 9. Although the transient process takes longer time, the fluctuation of the frequency is eliminated in all cases. This demonstrates the significant effect of communication network on the performance of the control system in microgrids. The optimal setting of PI controller therefore should take into account the delay from the communication and should be robust in case of latency modifications.

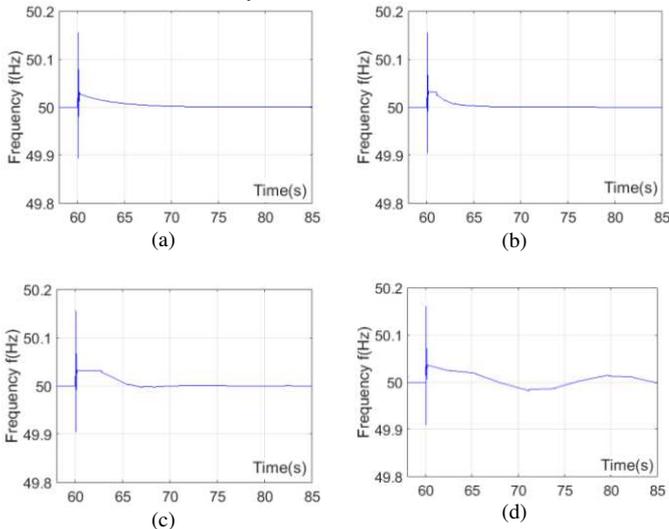

Figure 8. The frequency when tripping load 2
a) without communication network, b) with latency test case 1, c) with latency test case 2, d) with latency test case 3

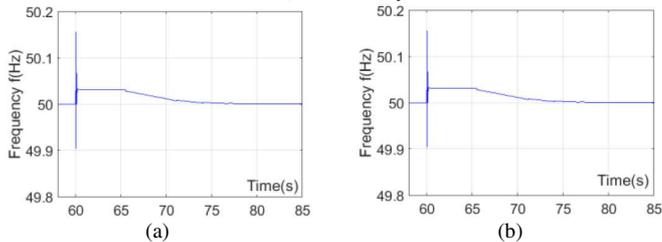

Figure 9. The frequency of the modified control system when tripping load 2
a) with latency test case 2, b) with latency test case 3

## V. CONCLUSION

This paper provided a holistic approach to validate the control system in microgrids with the consideration of the communication network. The communication model is developed in the FMI standard and integrated into the simulation of the microgrid. The two domains of power and communication could be studied simultaneously in the same simulator platform.

A test-case of microgrid with five inverter-based DGs experimented in both control approaches: centralization and distribution. A multi-agent platform is also built in Matlab/Simulink to implement the consensus algorithm in the distributed strategy. The results show that the operation of a microgrid is influenced by the performance of communication network. The evaluation of the stability of the control system hence needs the consideration of the data transmission.

The method used in this paper takes the advantages of FMI in the interoperability and reusability of the model. It could be used to study in many cases of the control system with various conditions of the communication network. In the future research, the designed FMU could be extended to take into account the impact of cyber- security and packet loss to the system or the evaluation the sensitivity of Power Hardware-in-the-loop testing with communication delay.

ACKNOWLEDGMENT

This work is supported by the H2020 Erigrid project, Grant Agreement No. 654113, with partial financial support from Vietnamese government.

The participation of G2Elab and CEA-INES is also partially supported by the Carnot Institute "Energies du Futur" under the PPInterop II project (www.energiesdufutur.eu).